\documentstyle{article}
\begin{document}
\renewcommand{\baselinestretch} {1.5}
\large

\vskip 1in
\begin{center}
{\large{\bf Quantum Monte Carlo Study of an Extended Hubbard Model  \\ 
on a Two-Leg Ladder}}
\vskip 0.5in
   Tomo MUNEHISA and Yasuko MUNEHISA\\
\vskip 0.3in
   Faculty of Engineering, Yamanashi University\\
   Kofu, Yamanashi, 400 \\
\end{center}
\vskip 0.8in
\noindent {\bf ABSTRACT}\\
\noindent
An extended Hubbard model on a two-leg ladder is numerically 
studied by means of the quantum Monte Carlo techniques. 
The model we study has the nearest-neighbor interactions which
are repulsive along chains and attractive for rungs.
The plots of the doping parameter versus the chemical potential 
show two cliff-like regions and a large plateau region.
Results on the charge susceptibility suggest it diverges in 
these cliff-like regions. These observations might imply a signal 
to the phase separation, which should be related to  
the effective attractive interactions along chains.

\eject
\noindent {\bf \S 1 \ Introduction}

The experimental realizations of the ladder systems\cite{r1},
which is considered as a useful bridge to clarify the 
relation between one-dimensional and two-dimensional systems,
have given theorists exciting topics in the study of the strongly 
correlated electrons.
Works on the quantum spin systems\cite{r2}, to which the half-filled 
Hubbard system reduces in the limit of the strong repulsive Coulomb 
interactions, show a striking contrast between the ladders with 
even legs, for which the finite spin gaps exist, and the odd-legged 
ladders which, as well as the one-dimensional (chain) case, show no 
spin gaps.
The $t$-$J$ model on ladders is another object to be studied\cite{r3}
in order to understand the doped system with the strong on-site Coulomb 
repulsion. 
Although we have not come to a very definite conclusion yet, 
interesting features like the $d$-wave correlations are suggested.
Furthermore some numerical studies for the Hubbard ladder model 
have been carried out\cite{r4}, but the issue of the 
long-range order of the pair-pair correlation functions seems  
controversial.

Recently various extended Hubbard models are under study to find 
the effects of the longer range interactions between the electrons\cite{r5}.
In this paper we report numerical results obtained by the quantum 
Monte Carlo techniques on an extended Hubbard model with the two-leg 
ladder, whose Hamiltonian 
includes the Coulomb-like nearest-neighbor interactions that are 
repulsive along chains and attractive for rungs,
studied in the conditions that the system is doped by holes or by
electrons and the on-site Coulomb repulsion lies in the intermediate
region.
It should be noted that, 
for the values of the model's parameters we use in the simulation,
the double occupancy is abundant 
so that holes can emerge even when electrons are doped to the system.
Our motivation to study this model is to investigate 
behaviors caused by double-occupancy-pairs or vacancy-pairs on rungs, 
where (and hereafter) we refer to the state of a vacant site, 
a site upon which no electrons are sitting, as vacancy.

We measure some thermodynamical quantities concerning to the number 
of particles of the system such as the doping parameter and the charge 
susceptibility as a function of the chemical potential. We observe 
that the plots of the doping parameter versus the chemical potential 
show {\it two unexpected cliff-like regions at finite dopings} 
in addition to a few plateau regions,
which have been expected from the results for the standard Hubbard model 
on the chain or the square lattice systems\cite{r6}.
The charge susceptibility measured in each of the cliff-like regions shows
a large peak which seemingly diverges at the temperature consistent with 
zero to imply the possibility of the phase separation that has not
been confirmed in the preceding studies of the Hubbard-like models.

In \S 2 we make a brief description on the model and method.
The Monte Carlo results are in \S 3 together with the definitions 
of the quantities we measured.
Finally \S 4 is devoted to discussions.

\vskip 0.5in
\noindent {\bf \S 2 \ the Model}

The Hamiltonian we study is
\begin{eqnarray}
\nonumber
   {\cal H} = -t_c
 \sum_{\sigma} \sum_{l} \sum_{i=1}^{N_r-1}
                       [c^\dagger_{i,l,\sigma}c_{i+1,l,\sigma} + h.c.]
             -t_r \sum_{\sigma} \sum_{i=1}^{N_r}
                       [c^\dagger_{i,a,\sigma}c_{i,b,\sigma} + h.c.] \\
             +V_c 
      \sum_{\sigma, \sigma'} \sum_{l} \sum_{i=1}^{N_r-1}
                        n_{i,l,\sigma}n_{i+1,l,\sigma'}
             +V_r \sum_{\sigma, \sigma'} \sum_{i=1}^{N_r}
                        n_{i,a,\sigma}n_{i,b,\sigma'}  
             +U   \sum_{l} \sum_{i=1}^{N_r}
                        n_{i,l,\uparrow}n_{i,l,\downarrow}
\label{eq:Hehb}
\end{eqnarray}
where $c_{i,l,\sigma}$ denotes the annihilation operator for an electron  
with spin $\sigma$ ($\uparrow$ or $\downarrow$)
which is located on the $i$-th rung along 
the leg $l$ ($a$ or $b$) of a ladder,
$N_r$ being the total number of rungs and 
$n_{i,l,\sigma} \equiv c^\dagger_{i,l,\sigma}c_{i,l,\sigma}$.
We employ an open boundary condition to avoid the hoppings 
between the sites on the first rung and the sites on the $N_r$-th rung.

The partition function $Z$ is given by 
\begin{eqnarray}
     Z={\rm tr} \lbrace e^{-\beta ({\cal H} - \mu{\cal N})} \rbrace , 
\label{eq:Z}
\end{eqnarray}
with the inverse temperature $\beta$, the chemical potential $\mu$
and 
\begin{eqnarray}
     {\cal N} \equiv 
     \sum_{\sigma} \sum_{l}  \sum_{i=1}^{N_r} n_{i,l,\sigma},
\label{eq:Ne}
\end{eqnarray}
which is the total number of electrons sitting on the ladder.

In the Suzuki-Trotter formula\cite{st} 
with the finite Trotter number $N_t$, 
\begin{eqnarray}
 Z  \simeq  \sum_{\alpha_1}\sum_{\alpha_2} \cdots \sum_{\alpha_{N_t}}
\langle \alpha_1 \mid e^{-\beta {\cal J}/{N_t}} 
\mid \alpha_2 \rangle \cdots \langle \alpha_{N_t}
\mid e^{-\beta {\cal J}/{N_t}} \mid \alpha_1 \rangle ,
\label{eq:Znt}
\end{eqnarray}
where we denote ${\cal H}-\mu {\cal N}$ by $\cal J$.
We employ a complete set \cite{mune}
\begin{eqnarray}
\mid \alpha \rangle = \mid S_1, S_2, \cdots S_{N_r} \rangle
\label{eq:cpls}
\end{eqnarray}
with sixteen states in  Table I to denote the state 
$S_i$ on the $i$-th rung, which are the eigenstates for  
\begin{eqnarray}
\nonumber
j_i   
\equiv  -\frac{t_r}{2} \sum_{\sigma} 
                 [c^\dagger_{i,a,\sigma}c_{i,b,\sigma} + h.c.]
+\frac{V_r}{2} \sum_{\sigma, \sigma'} n_{i,a,\sigma}n_{i,b,\sigma'} 
    +\frac{U}{2} \sum_{l} n_{i,l,\uparrow}n_{i,l,\downarrow} \\
-\frac{\mu}{2} \sum_{\sigma, \sigma'}
    \sum_{l} n_{i,l,\sigma}n_{i,l,\sigma'},
\label{eq:ji}
\end{eqnarray}
so that the checkerboard decomposition is available for
${\cal J} ={\cal J}_1 +{\cal J}_2$,
\begin{eqnarray}
\nonumber
{\cal J}_1  \equiv  
    -t_c \sum_{\sigma} \sum_{l} \sum_{k=1}^{N_r/2}
                       [c^\dagger_{2k-1,l,\sigma}c_{2k,l,\sigma} + h.c.]
    +V_c \sum_{\sigma, \sigma'} \sum_{l} \sum_{k=1}^{N_r/2}
                        n_{2k-1,l,\sigma}n_{2k,l,\sigma'} 
    +\sum_{i=1}^{N_r} j_i ,\\
\nonumber
{\cal J}_2  \equiv  
    -t_c \sum_{\sigma} \sum_{l} \sum_{k=1}^{N_r/2-1}
                       [c^\dagger_{2k,l,\sigma}c_{2k+1,l,\sigma} + h.c.]
    +V_c \sum_{\sigma, \sigma'} \sum_{l} \sum_{k=1}^{N_r/2-1}
                        n_{2k,l,\sigma}n_{2k+1,l,\sigma'} \\
    +\sum_{i=1}^{N_r} j_i ,
\label{eq:h1h2}
\end{eqnarray}
where an even $N_r$ is assumed.
Parameters $u_1$ and $u_2$ in Table I are given by 
\begin{eqnarray}
\nonumber
u_1 \equiv \frac{1}{2} \sqrt{1+ \frac{U}{\sqrt{U^2+16t_r^2}}}, \ \ \
\nonumber
u_2 \equiv \frac{1}{2} \sqrt{1- \frac{U}{\sqrt{U^2+16t_r^2}}}.
\label{eq:u1u2}
\end{eqnarray}

\vskip 0.5in
\noindent {\bf \S 3 \ Numerical results}

Now let us show our quantum Monte Carlo results on the 
model described in the previous section.
Here we concentrate our attention on particular values of 
parameters in the Hamiltonian (\ref{eq:Hehb}), 
$t_c=1$, $t_r=2$, $V_c=2$, $V_r=-4$ and $U=4$. For this choice 
it is possible to obtain statistically meaningful results
while the sign problem is quite serious for other choices we tried.

Because of the emergence of negative 
weights, we should measure a physical quantity $A$ in each thermally
equilibrated configuration together with its sign and obtain its
expectation value by
\begin{eqnarray}
\langle A \rangle_{\rm av} = {A_{net} \over Z_{net}}
 = {{A_+ - A_-} \over {Z_+ - Z_-}},
\label{eq:A}
\end{eqnarray}
where $Z_+ (Z_-)$ is the number of configurations with positive (negative)
weight, and $A_+ (A_-)$ denotes the contribution from positively
(negatively) signed configurations.
The $r$ ratio defined by 
\begin{eqnarray}
r \equiv  {{Z_+ - Z_-} \over {Z_+ + Z_-}},
\label{eq:r}
\end{eqnarray}
will inform us how serious the cancellation is.

Changing the number of rungs $N_r$ ($8 \le N_r \le 32$), 
the inverse temperature $\beta$ ($3.3 \le \beta \le 10$) 
and the chemical potential $\mu$, we measure the doping parameter 
\begin{eqnarray}
\delta \equiv 1- \frac{N_e}{N_s},
\label{eq:dlt}
\end{eqnarray}
where $N_e \equiv \langle {\cal N} \rangle_{\rm av}$ is the number of electrons 
and $N_s \equiv 2N_r$ the number of sites on the ladder,
and the charge susceptibility per site,
\begin{eqnarray}
\chi_c \equiv \frac{1}{N_s}
\frac{\partial N_e }{\partial \mu} = \frac{\beta}{N_s}
\,(\langle {\cal N}^2 \rangle_{\rm av} - \langle {\cal N} \rangle_{\rm av}^2).
\label{eq:chic}
\end{eqnarray}
Typically we generate 80000 configurations in one measurement 
by the local update and the global update in the Trotter direction, 
details of which are in \cite{update}, 
and last 60000 configurations are used for the  measurement.
Each datum is an average of six such measurements with different random 
number sequences and the statistical error of the datum is calculated 
from the standard deviation among these six runs.
As we observed very little discrepancy between the $N_t=32$ results and 
the $N_t=64$ ones we fixed $N_t$=64 throughout the measurements.
The $r$-ratio in the simulations varies from $0.17$ to $0.78$, depending 
on values of $N_r$, $\mu$ and $\beta$.

Figure~1 presents the doping parameter $\delta$ on $N_s=16$ (namely
$N_r=8$) and $N_s=64$ ladders at $\beta = 10$ 
as a function of the chemical potential $\mu$. 
We observe rapid increases of $\delta$ around $\mu \sim -1$, where 
the system is under the hole doping, and $\mu \sim 5$ which belongs 
to the region of the electron doping, together with three plateaus 
in between. 
Comparing the results for $N_s=16$ and $N_s=64$ ladders, we conclude
that these plateaus would be split due to the finite size effect and 
would merge to form the $\delta=0$ plateau in the thermodynamic limit,
implying a large charge gap $\Delta_c \sim 6$. 
It seems that $\delta$ changes smoothly in the neighborhood of the 
half-filling. 

Let us then present the data on $\chi_c$, 
the charge susceptibility per site, in more detail.
Figure~2(a) plots $\chi_c$ per site versus $\delta$ at $\beta=10$
on up to an $N_s=64$ ladder, choosing several values of $\mu$ with 
which we observe two outstanding enhancements of $\chi_c$
corresponding to the results in Fig.~1.
It is notable that for each value of $\mu$ in the negative 
$\delta$ region the values of $\chi_c$ and $\delta$ 
strongly depend on the size of the ladder, 
while the data for positive $\delta$ are much less size-dependent 
except for the value of $\chi_c$ around $\delta \sim 0.2$.
In both the positive and the negative $\delta$ regions 
the thermodynamic limit of the peak value of $\chi_c$ 
for each value of $\beta$, which we denote by $\chi_c^m$, is 
successfully obtained by the linear extrapolation of the data 
$1/\chi_c$ at the peak for each ladder size 
versus $1/N_s$ with $N_s =16,$ 24, 32, 48 and 64.
In Fig.~2(b) we plot these extrapolated values
as a function of the temperature $T \equiv 1/\beta$.
(Note that the ordinate measures the inverse of $\chi_c^m$.)
Linearly fitted lines to the data drawn in the figure suggests
that $1/\chi_c^m$ reduces to zero (namely $\chi_c^m$ goes to infinity) 
near $T = 0$ in both regions of $\delta$.
Figure~2(c) shows the values of $\delta$ to maximize $\chi_c$ 
evaluated from the data in Fig.~2(a), which we denote by $\delta_0$,  
as a function of the $1/N_s$ at temperatures $T=0.1$, $0.2$ and $0.3$. 
Because of large uncertainties compared to those on the $\chi_c$ at 
peaks it is difficult to estimate reliable critical value  
$\delta_c$ at which the charge susceptibility diverges, but it is 
intriguing to note the data suggest finite critical values 
$\delta_c \sim 0.2$ and $\delta_c \sim -0.2$.
   
\vskip 0.5in
\noindent {\bf \S 4 \ Discussions}

In the previous section we showed our quantum Monte Carlo results 
on the extended Hubbard ladder model described in \S 2. 
What we measure in the simulation are the quantities related to the 
particle density, the doping parameter $\delta$ and the 
charge susceptibility per site $\chi_c$,
which could signal the charge gap and the phase separation.
We find two regions where $\chi_c$ is greatly enhanced, 
one of which corresponds to the electron doping ($\delta < 0$) 
and the other to the hole doping ($\delta > 0$). 
Linear extrapolations on these enhancements suggest that 
$\chi_c$ diverges for the dopings away from the half-filling
at the temperatures which are consistent with zero.
Between these singularities we observe three plateaus which would 
merge to one in the thermodynamic limit to indicate 
the existence of the large charge gap.

We observe no sign for the singularity of 
$\chi_c$ near $\delta=0$, which is in marked contrast to the 
observations in the one- and two-dimensional standard Hubbard cases
where $\chi_c$ behaves like $1/\delta$ when $\delta \sim 0$\cite{r6,xic}. 
Since it seems difficult to understand our results in the same manner 
as the latter cases,
this feature, together with the enhancement of $\chi_c$ at finite 
values of $\delta$, implies that a new picture is  
necessary to explain the behaviors of our model.
We would like to stress that this is not quite surprising because 
studies of the spin $1/2$ ladder systems, 
which corresponds to the $U \rightarrow \infty$ limit of the half-filled 
Hubbard models on a ladder, present a picture that differs much from 
those for the spin $1/2$ models on a chain or on a plane\cite{r1,r2}.

What happens to the model in these regions with the increasing 
$\chi_c$ around the non-zero $\delta_c$ then? 
Although it is an open question so far 
and much more work would be necessary to answer it, one clue might 
be found in the behavior of the
double-occupancy-pairs on rungs in the positive $\delta$ region 
as well as the vacancy-pairs on rungs in the negative $\delta$ region. 

Let us comment on the hole doping case first. Here we examined  
the number of doubly occupied sites $N_d$ and 
the number of double-occupancy-pairs on rungs $N_{dd}$ defined by
\begin{eqnarray}
N_d \equiv \langle {\cal N}_{d} \rangle_{\rm av},  \ \ \
{\cal N}_{d} \equiv \sum_{l=a,b}\sum_{i=1}^{N_r} 
n_{i,l,\uparrow} n_{i,l,\downarrow} 
\label{eq:numd}
\end{eqnarray}
\begin{eqnarray}
N_{dd} \equiv \langle {\cal N}_{dd} \rangle_{\rm av}, \ \ \  
{\cal N}_{dd} \equiv \sum_{i=1}^{N_r}
n_{i,a,\uparrow} n_{i,a,\downarrow}
n_{i,b,\uparrow} n_{i,b,\downarrow},
\label{eq:numdd}
\end{eqnarray}
respectively.  In the measurement of these quantities  
for the hole-doped $N_s=64$ ladders at $\beta = 10$, 
we observe that the ratio of doubly occupied sites to all sites, 
$N_d/N_s$, linearly increases from 0 to $\sim $0.4 as $\delta$ decreases 
from $\sim 0.5$ to $\sim 0$. Comparing this ratio with the number of 
double-occupancy-pairs on rungs divided by the number of rungs, 
$N_{dd}/N_r$, we see that most of these doubly occupied sites prefer 
pairing on rungs.
In order to clarify contributions of the double-occupancy-pairs on rungs
to the charge susceptibility $\chi_c$, we then divide $\chi_c$ into three 
parts $P_1$, $P_2$ and $P_3$ and measured them separately. 
Here we let $P_1$ represent the portion  
which is proportional to the susceptibility concerning to the 
double-occupancy-pairs on rungs, $\chi_{dd}$,    
\begin{eqnarray}
 P_1 \equiv \frac{\beta}{N_s}(\langle {4{\cal N}_{dd}}^2 \rangle_{\rm av} 
- \langle {4{\cal N}_{dd}} \rangle_{\rm av}^2) 
= \beta \times 8 \chi_{dd}, 
\label{eq:p1}
\end{eqnarray}
\begin{eqnarray}
 \chi_{dd} \equiv \frac{1}{N_r}(\langle {{\cal N}_{dd}}^2 \rangle_{\rm av} 
- \langle {{\cal N}_{dd}} \rangle_{\rm av}^2), 
\label{eq:chidd}
\end{eqnarray}
$P_2$ the portion concerning to other states,
\begin{eqnarray}
 P_2 \equiv \frac{\beta}{N_s}(\langle {{\cal N}_{oth}}^2 \rangle_{\rm av}
- \langle {{\cal N}_{oth}} \rangle_{\rm av}^2), \ \ \
{\cal N}_{oth} \equiv {\cal N}-4{\cal N}_{dd},
\label{eq:p2}
\end{eqnarray}
and $P_3$ the interference term 
\begin{eqnarray}
\nonumber
 P_3 \equiv 2\frac{\beta}{N_s}
(\langle {{\cal N}_{dd}}{{\cal N}_{oth}} \rangle_{\rm av}
- \langle {{\cal N}_{dd}} \rangle_{\rm av} 
 \langle {{\cal N}_{oth}} \rangle_{\rm av} ).
\label{eq:p3}
\end{eqnarray}
Figure~3 shows results for $P_1$ and $P_2$ on an $N_s=64$ 
ladder with $\beta=10$ together with $\chi_c$ as a function of the 
chemical potential $\mu$. We clearly see that the double-occupancy-pairs 
on rungs have dominant effects in the rapid increase of $\chi_c$.

In the case of the electron doping we turn our attention to 
the number of the vacant sites, $N_v$, calculated by 
\begin{eqnarray}
N_v = N_s - N_d -(N_e -2 N_d) = N_s \delta + N_d, 
\label{eq:numv}
\end{eqnarray}
the number of the vacancy-pairs on rungs, $N_{vv}$, and 
the susceptibility concerning to these pairs, $\chi_{vv}$, which are
defined using $h_{i,l,\sigma} \equiv 1-n_{i,l,\sigma}$
instead of $n_{i,l,\sigma}$ in eqs.(\ref{eq:numdd}) and (\ref{eq:chidd}).
We observe approximately particle-hole symmetric behaviors of $N_v$ and
$N_{vv}$ for $-0.5 < \delta < -0.1$ in comparison with the hole 
doping case. 
We also observe that the susceptibility $\chi_{vv}$ shows a
remarkable enhancement which suggests that the vacancy-pairs play an 
important role when $\chi_c$ diverges in the region $\delta \sim -0.3$, 
although the relation between $\chi_{vv}$ and $\chi_c$ is more 
complicated than that between $\chi_{dd}$ and $\chi_c$.   

Several remarks on the future work are in order.
Correlation functions corresponding to the susceptibilities 
$\chi_c$, $\chi_{dd}$ and $\chi_{vv}$ would be worth studying
to understand whether the long-range order exists in this model.
It will be important to measure the magnetic susceptibility and 
the specific heat in order to shed the light on the physics   
of the model.
As was mentioned in \S 2, we had to limit ourselves to a particular 
choice of parameters $t_c$, $t_r$, $V_c$, $V_r$ and $U$ in the simulation 
because of the sign problem. It would be a future task to study this model 
for other values of the parameters.
It would be also interesting to investigate such an extended model in a 
bilayer system whose planes are the two-dimensional counterparts of 
the chains in the ladder system, where the realistic
choice of parameters might be realized.

\eject

%\eject
\vskip 0.5in
\noindent {\bf Figure Captions}

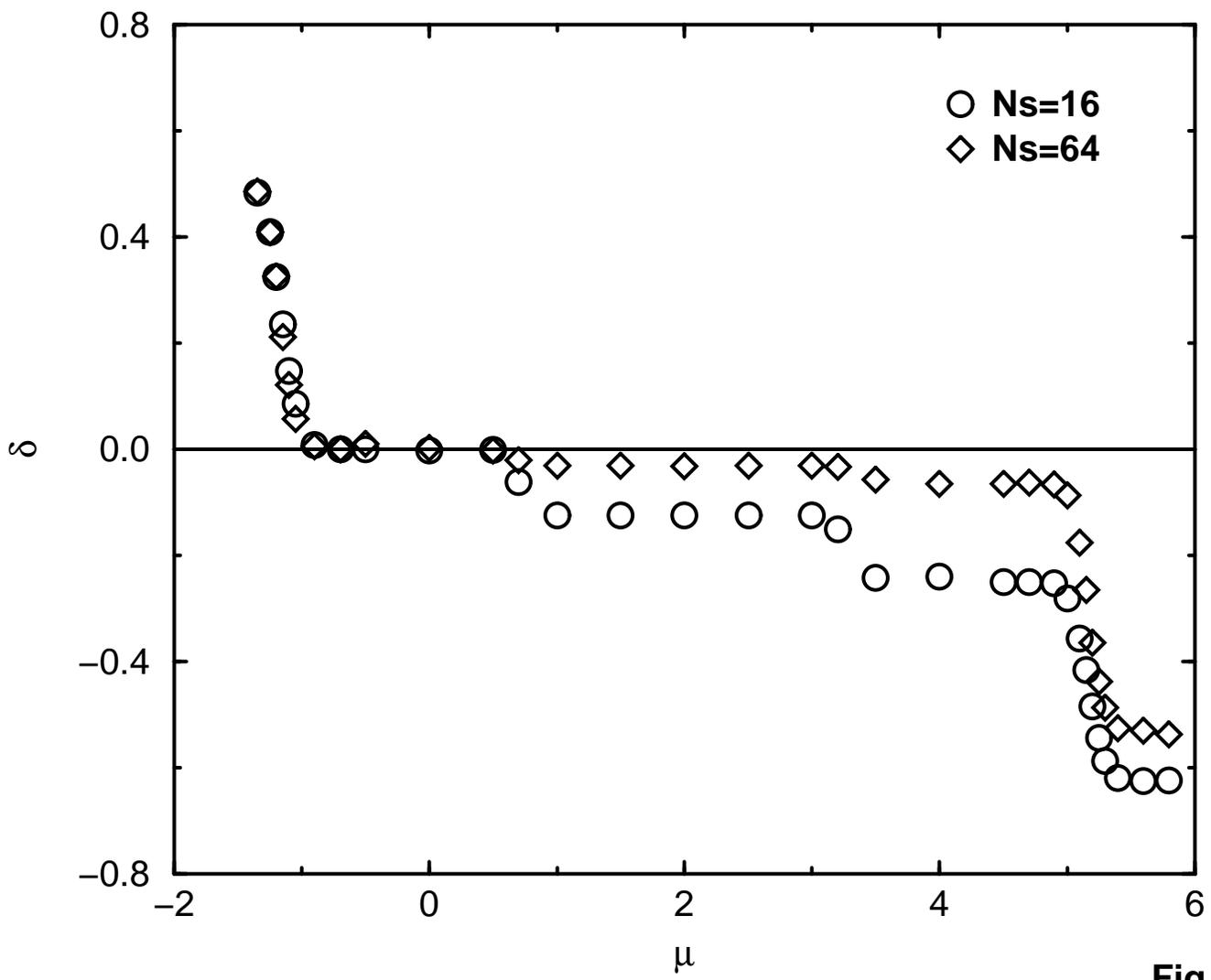
\noindent {\bf Figure \ 1} \\
Monte Carlo results on the doping parameter $\delta$ 
as a function of the chemical potential $\mu$ measured on $N_s = 16$ 
and $N_s=64$ ladders at the inverse temperature $\beta = 10$.
All the statistical errors are within symbols.

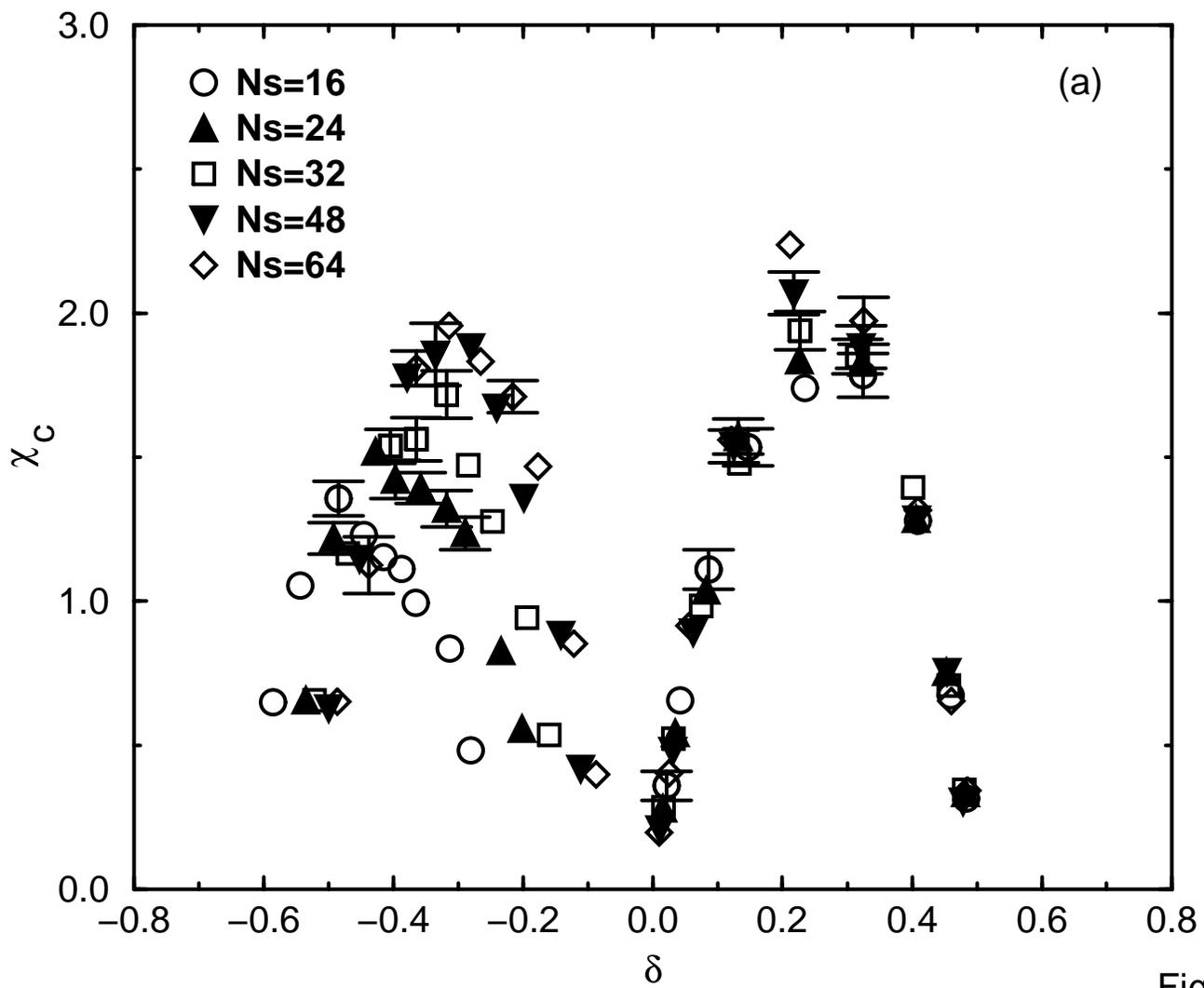
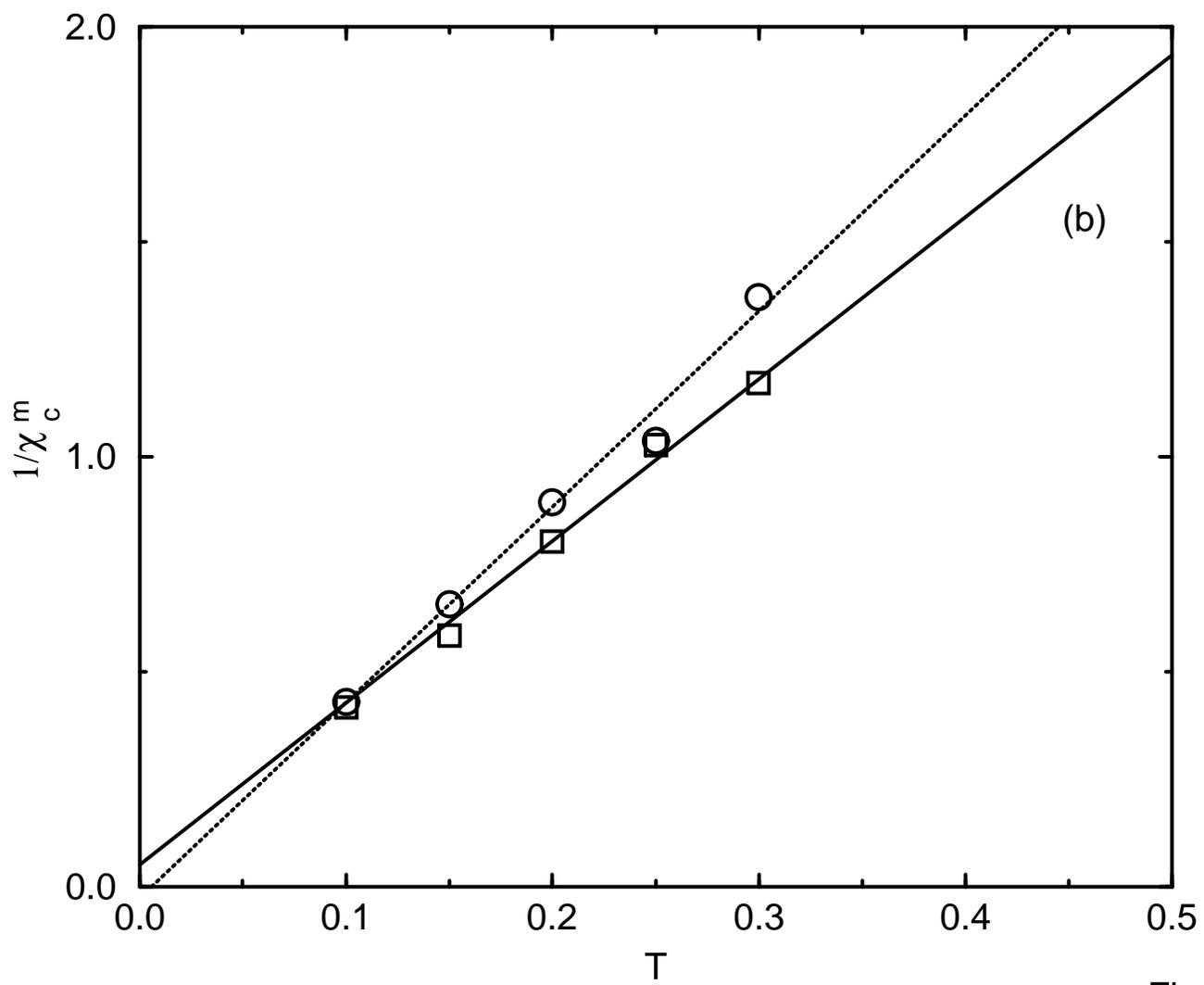
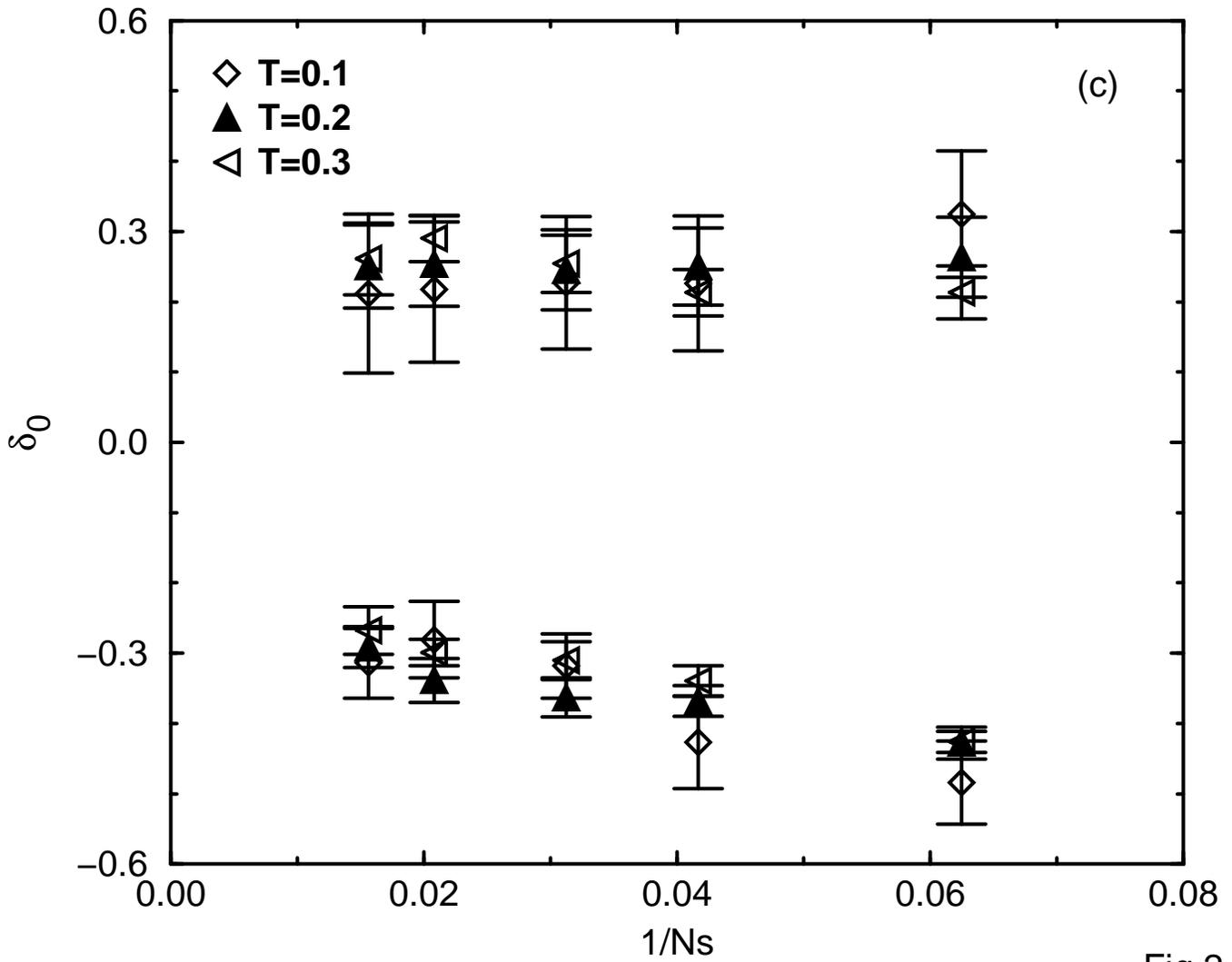
\noindent {\bf Figure \ 2} \\
\noindent {\bf (a)} 
The charge susceptibility per site, $\chi_c$,
versus the doping parameter $\delta$ 
measured for the chemical potential ranging
$-1.35 \le \mu \le -0.95$ and $5.0 \le \mu \le 5.3$
on the $N_s=$16, 24, 32, 48 and 64 ladders at $\beta=10$.
All the statistical errors for $\delta$ are within symbols.
The statistical errors for $\chi_c$ are within symbols if not 
shown explicitly.

\noindent {\bf (b)} 
Temperature dependence of the inverse 
peak value of the charge susceptibility per site extrapolated to 
$N_s \rightarrow \infty$, which we denote by $1/\chi_c^m$.
Squares (circles) are symbols for the data with negative (positive) $\delta$.
Errors evaluated in the process of the extrapolation are within symbols.
The result from the linear fit for the data with negative (positive) 
$\delta$ is also shown in the figure with a solid (dotted) line.

\noindent {\bf (c)} 
Values of $\delta_0$, the values of $\delta$ to maximize 
$\chi_c$, at $T=0.1$, $0.2$ and $0.3$ as a function of $1/N_s$.
We evaluate $\delta_0$ by the peak among the measured data in 
Fig.~2(a) and estimate its errors by the difference of
the two values of $\delta$ around the peak. 

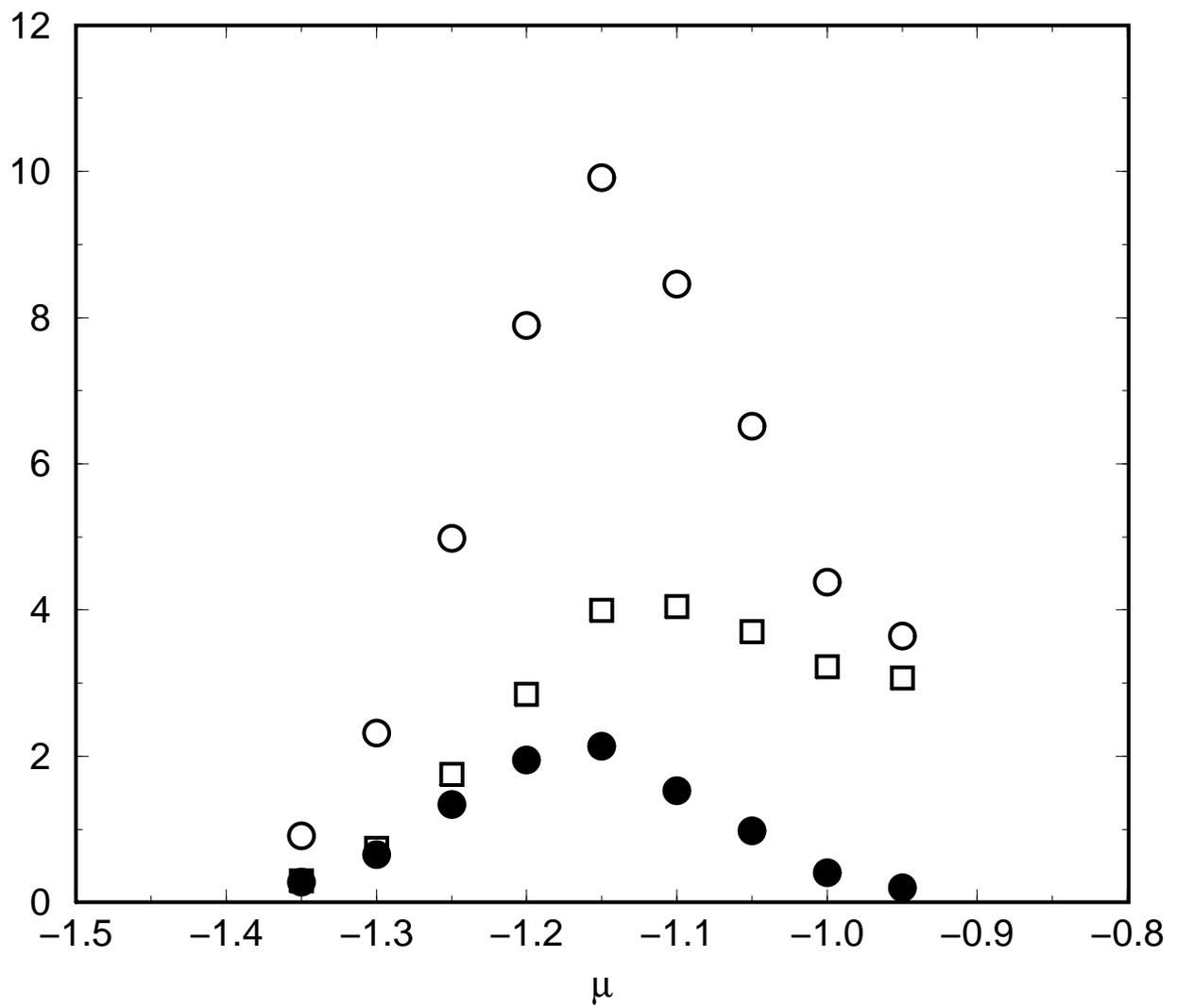
\noindent {\bf Figure \ 3} \\
Contributions of the double-occupancy-pairs on rungs and other states 
to the charge suseptibility $\chi_c$
measured in the positive $\delta$ region on an $N_s=64$ ladder 
at the inverse temperature $\beta=10$, 
as a function of the chemical potential $\mu$. 
Open circles, open squares and filled circles represent $P_1$, $P_2$ and 
$\chi_c$ defined by eqs. (\ref{eq:p1}),(\ref{eq:p2}) and (\ref{eq:chic}),
respectively.  Statistical errors are within symbols.

\eject
\noindent {\bf Table Captions} \\
\noindent {\bf Table \ I}\\
States on two sites of the $i$-th rung
         used to construct a complete set in
         the Suzuki-Trotter formula. $\mid 00 \rangle$
         represents a state with no electron on either site
         of the rung.

\vskip 0.5in

%\begin{table}[htbp]
%\begin{center}
\begin{tabular}{|r|c|}  \hline
  $ No. $ &  state \\ \hline
  $1$ & $\mid 00 \rangle $ \\ \hline
  $2$ & $\frac{1}{\sqrt{2}}(c^\dagger_{i,a,\uparrow}
   +c^\dagger_{i,b,\uparrow})\mid 00 \rangle $ \\ \hline
  $3$ & $\frac{1}{\sqrt{2}}(c^\dagger_{i,a,\uparrow}
   -c^\dagger_{i,b,\uparrow})\mid 00 \rangle $ \\ \hline
  $4$ & $\frac{1}{\sqrt{2}}(c^\dagger_{i,a,\downarrow}
   +c^\dagger_{i,b,\downarrow})\mid 00 \rangle $ \\ \hline
  $5$ & $\frac{1}{\sqrt{2}}(c^\dagger_{i,a,\downarrow}
   -c^\dagger_{i,b,\downarrow})\mid 00 \rangle $ \\ \hline
  $6$ & $c^\dagger_{i,a,\uparrow}
         c^\dagger_{i,b,\uparrow}\mid 00 \rangle $ \\ \hline
  $7$ & $\frac{1}{\sqrt{2}}(c^\dagger_{i,a,\uparrow}
                            c^\dagger_{i,b,\downarrow}
                           +c^\dagger_{i,a,\downarrow}
         c^\dagger_{i,b,\uparrow}) \mid 00 \rangle $ \\ \hline
  $8$ & $\frac{1}{\sqrt{2}}(c^\dagger_{i,a,\uparrow}
                            c^\dagger_{i,a,\downarrow}
                          - c^\dagger_{i,b,\uparrow}
                   c^\dagger_{i,b,\downarrow}) \mid 00 \rangle $ \\ \hline
  $9$ & $[u_1 (c^\dagger_{i,a,\uparrow}
                            c^\dagger_{i,b,\downarrow}
         -c^\dagger_{i,a,\downarrow}
                           c^\dagger_{i,b,\uparrow})
        +u_2 (c^\dagger_{i,a,\uparrow}
                            c^\dagger_{i,a,\downarrow}
        + c^\dagger_{i,b,\uparrow}
                            c^\dagger_{i,b,\downarrow})]
\mid 00 \rangle $ \\ \hline
  $10$ & $[u_2 (c^\dagger_{i,a,\uparrow}
                            c^\dagger_{i,b,\downarrow}
         -c^\dagger_{i,a,\downarrow}
                           c^\dagger_{i,b,\uparrow})
        -u_1 (c^\dagger_{i,a,\uparrow}
                            c^\dagger_{i,a,\downarrow}
        + c^\dagger_{i,b,\uparrow}
                            c^\dagger_{i,b,\downarrow})]
\mid 00 \rangle $ \\ \hline
  $11$ & $c^\dagger_{i,a,\downarrow}
         c^\dagger_{i,b,\downarrow}\mid 00 \rangle $ \\ \hline
  $12$ & $\frac{1}{\sqrt{2}}
   (c^\dagger_{i,a,\uparrow} c^\dagger_{i,a,\downarrow}
    c^\dagger_{i,b,\uparrow}
   +c^\dagger_{i,a,\uparrow}
    c^\dagger_{i,b,\uparrow} c^\dagger_{i,b,\downarrow})
   \mid 00 \rangle $ \\ \hline
  $13$ & $\frac{1}{\sqrt{2}}
   (c^\dagger_{i,a,\uparrow} c^\dagger_{i,a,\downarrow}
    c^\dagger_{i,b,\uparrow}
   -c^\dagger_{i,a,\uparrow}
    c^\dagger_{i,b,\uparrow} c^\dagger_{i,b,\downarrow})
   \mid 00 \rangle $ \\ \hline
  $14$ & $\frac{1}{\sqrt{2}}
   (c^\dagger_{i,a,\uparrow} c^\dagger_{i,a,\downarrow}
    c^\dagger_{i,b,\downarrow}
   +c^\dagger_{i,a,\downarrow}
    c^\dagger_{i,b,\uparrow} c^\dagger_{i,b,\downarrow})
   \mid 00 \rangle $ \\ \hline
  $15$ & $\frac{1}{\sqrt{2}}
   (c^\dagger_{i,a,\uparrow} c^\dagger_{i,a,\downarrow}
    c^\dagger_{i,b,\downarrow}
   -c^\dagger_{i,a,\downarrow}
    c^\dagger_{i,b,\uparrow} c^\dagger_{i,b,\downarrow})
   \mid 00 \rangle $ \\ \hline
  $16$ & $ c^\dagger_{i,a,\uparrow} c^\dagger_{i,a,\downarrow}
           c^\dagger_{i,b,\uparrow} c^\dagger_{i,b,\downarrow}
   \mid 00 \rangle $ \\ \hline
\end{tabular}

\vskip 0.1in

\rightline{  Table I}
%\end{center}
%\end{table}
%%\eject
%%\begin{figure}[p]
%%\centering
%%\epsfile{file=fig1.eps,scale=0.7}
%%\epsfile{file=fig2a.eps,scale=0.7}
%%\end{figure}

%%\eject
%%\begin{figure}[p]
%%\centering
%%\epsfile{file=fig2b.eps,scale=0.7}
%%\epsfile{file=fig2c.eps,scale=0.7}
%%\end{figure}

%%\eject
%%\begin{figure}[p]
%%\centering
%%\epsfile{file=fig3.eps,scale=0.7}
%%\end{figure}

\end{document}